\RequirePackage{lineno}
\documentclass[preprint,showpacs,showkeys,preprintnumbers,amsmath,amssymb,lineno]{revtex4}

\usepackage{graphicx}
\usepackage{graphics}
\usepackage{dcolumn}
\usepackage{bm}

\begin{document}
\preprint{}
\title{Density of states in solid deuterium: Inelastic neutron scattering study
\\}

\author{ A.\ Frei, E.\ Gutsmiedl$^{*}$, C.\ Morkel, A.R.\ M\"uller, S.\ Paul, M.\ Urban}

\affiliation{Technische Universit\"at M\"unchen,
 Physik-Department,
 James-Franck-Str., D-85747 Garching, Germany}

\author{H.\ Schober, S.\ Rols}
\affiliation{Institut Laue Langevin, 156X, F-38042 Grenoble CEDEX,
France}

\author{T.\ Unruh}
\affiliation{Technische Universit\"at
M\"unchen,Forschungsneutronenquelle Heinz Maier-Leibnitz (FRM II),
Lichtenbergstr. 1, D-85747 Garching, Germany}

\author{M.\ H\"olzel}
\affiliation{Technische Universit\"at Darmstadt, Material-und
Geowissenschaften, Petersenstr. 23, D-64287 Darmstadt, Germany}

\date{\today}

\begin{abstract}

 The dynamics of solid deuterium (sD$_2$) is
studied by means of inelastic scattering (coherent and incoherent)
of thermal and cold neutrons at different temperatures and
para-ortho ratios. In this paper, the results for the generalized
density of states (GDOS) are presented and discussed. The
measurements were performed at the thermal neutron time-of-flight
(TOF) instrument IN4 at ILL Grenoble and at the cold neutron TOF
instrument TOFTOF at FRM II Garching. The GDOS comprises besides
the hcp phonon excitations of the sD$_2$ the rotational
transitions $J=0 \mapsto 1$ and $J=1\mapsto 2$. The intensities of
these rotational excitations
 depend strongly on the ortho-D$_2$ molecule concentration $c_o$ in sD$_2$.
Above $E=10$~meV there are still strong excitations, which very
likely may originate from higher energy damped optical phonons and
multi-phonon contributions. A method for separating the one- and
multi-phononon contributions to the density of states will be presented and discussed.\\

\end{abstract}

\pacs{28.20.Cz, 63.20.kk}
\keywords{neutron, phonons, solid deuterium}

\noindent
$^*$Corresponding author; email: egutsmie@e18.physik.tu-muenchen.de\\

\maketitle

\section{\label{sec:level1} Introduction}
Solid deuterium (sD$_2$) and solid hydrogen are typical quantum
molecular solids. Each D$_2$/H$_2$ molecule exhibits large
zero-point vibrations due to the small molecule mass. These
quantum solids have been investigated theoretically \cite{MDC1}
and by experimental techniques like inelastic neutron scattering
\cite{Nielsen1} and Raman scattering \cite{Dries}. Solid deuterium
has an almost perfect hcp crystal structure, when it is prepared
under suitable conditions \cite{Silv} (low pressure and $T >
5$~K). The phonon dispersion relation of solid ortho-deuterium
(o-D$_2$) was measured in the past by inelastic coherent neutron
scattering \cite{Nielsen2}. These measured phonon dispersion
relations are in good agreement with recent calculations based on
molecular dynamics calculations (MDC) \cite{MDC1}. In the case of
solid para-hydrogen, the molecular calculations \cite{MDC1}
disagree to some extent with the experimental data of
\textit{Nielsen} \cite{Nielsen1}. The MDC predicts phonons above
$10$~meV in solid para-hydrogen, which did not appear in the
neutron scattering measurements of \textit{Nielsen}
 \cite{Nielsen1}. One possible explanation could be the restriction
to energies below $9$~meV in Nielsen's data analysis \cite{MDC1}.
Recent inelastic incoherent neutron scattering data on solid
para-hydrogen
\cite{Ishmaev} are showing also excitations above $10$~meV.\\
\\
The D$_2$ molecule has internal rotational modes, which are
described by the rotational quantum number $J$. In the solid
phase, $J$ is still a good quantum number. Deuterium in states
with even $J$ ($J$=0,2,4....) is termed ortho-deuterium (o-D$_2$),
whereas D$_2$ in states with odd $J$ ($J$=1,3,5,.....) is termed
para-deuterium (p-D$_2$).
 At low temperatures $(T\sim 6$~K), about 99.999$\%$ of the deuterium
molecules are in the ortho state, when thermal equilibrium is
reached. At room temperature, D$_2$ has an ortho concentration of
$66.7\%$. After cooling down the D$_2$ to the solid phase  $(
T\sim 6$~K), it normally takes months to reach the equilibrium of
$99.999\%$ o-D$_2$. This process can be accelerated by using
paramagnetic materials like chromium oxides or hydrous ferric
oxides \cite{PO1,PO2,PO3} as catalysts for the para-ortho
conversion. With this kind of converters, it is possible to reduce
the conversion time to below one day in typical experiments and to
adjust the o-D$_2$ concentration between $66.7\%$ and $>=98\%$ for
investigating in detail its influence on the structure and dynamics of solid deuterium.\\
\\
Neutron scattering is an excellent tool to investigate the phonon
system and the rotational transitions of D$_2$ molecules in
 sD$_2$. Neutrons are scattered by D coherently
($b_{\mathrm{coh}}=6.671$~fm) and also incoherently
($b_{\mathrm{inc}}=4.04$~fm) \cite{NSL}. Scattering of neutrons
where a rotational transition ($J \mapsto J'$, $J\neq J'$) is
involved, leads to an incoherent response of the system. The
phonon excitations are present in the coherent part of the
scattering, whereas the incoherent
 part is determined by rotational transitions and also by
incoherent phonon scattering \cite{Lov,Liu2}.
\\
In principle the cross section for neutron scattering in sD$_2$ is
described by,

\begin{equation}
\label{eq:csall}
 \frac{\partial^{2}\sigma}{\partial {\omega}
\partial {\Omega}}={\frac {k}{{\it k_0}} {{\it
b_{\mathrm{coh}}}}^{2}S_{\mathrm{coh}}\left(Q,\omega \right) }+
{\frac {k}{{\it k_0}} {{\it
b_{\mathrm{inc}}}}^{2}S_{\mathrm{inc}}\left(Q,\omega \right) }.
\end{equation}
\\
Here, $k_0$ is the wave number of the incoming neutron, k the wave
vector of the scattered neutron, $S_{i} \left( Q,\omega \right)$
($i=$coh, inc) are the dynamical structure functions of coherent
and incoherent
 scattering, respectively. The energy change of the neutron is $E=\hbar \omega$ ,
while $\hbar Q$ is the momentum transfer. The coherent and
incoherent
scattering lengths are $b_\mathrm{coh}$ and $b_\mathrm{inc}$~respectiveley.  \\
\\
In the incoherent approximation the neutron scattering cross
section Eq.\ \eqref{eq:csall} can  be expressed by \cite{Squires}:

\begin{equation}
\label{eq:csinc}
\frac{\partial^{2}\sigma}{\partial\omega\partial\Omega}=\frac{k}{k_0}
\left[b_{\rm{eff}}(Q)\right]^{2}\frac{\hbar
Q^{2}}{2M}\frac{G(\omega)}{\omega}\left[n(\omega)+1\right]e^{-2W(Q)}.
\end{equation}

The generalized density of states (GDOS) $G(\omega)$ comprises the
complete phonon excitations of sD$_2$ as well as rotational
transitions of individual  D$_2$ molecules. Furthermore,
 multi-phonon excitations of the phonon system of sD$_2$
\cite{Mertens} should appear in the GDOS, as they are not
corrected for in Eq.\ \eqref{eq:csinc}. The term $e^{-2W(Q)}$ is
the well known Debye-Waller factor, where $W(Q)=\frac{1}{6}
Q^{2}\langle u^{2} \rangle$,  and $\langle u^{2} \rangle$
($\langle u^{2} \rangle =0.25$~\AA$^{2}$) is the mean square
displacement \cite{Nielsen2} of the D$_2$ molecule in the lattice.
The quantity $n(\omega)$ is the Bose statistic function, and
$[n(\omega)+1]$ describes the creation of a boson, which causes
the neutron energy loss. The scattering length $b_{\rm{eff}}(Q)$
is a combination of coherent and incoherent scattering lengths.
This combination depends strongly on the rotational transitions excited by the neutrons \cite{Liu2} (see TABLE I). \\
\\
Factorization (Eq.\ \eqref{eq:csinc}) is only applicable, if the
total scattering response of solid deuterium contains an essential
amount of incoherent scattering (incoherent approximation), or if
averaging over a sufficiently large Q range is performed. This is
the case for thermal neutron scattering.
\\
\\

\begin{table}[h]
\caption{\label{tab:table1} Neutron scattering lengths $b$
 associated with rotational transitions \cite{Liu2}.}
\begin{tabular}{|l|c|l|}
\hline {\bf $J \mapsto\tilde{J}$}\rule[0pt]{0pt}{2.8ex} & { $E$
[meV]} & \bf $b^{2}$
\\
\hline\hline
\rule[0pt]{0pt}{2.8ex}    $0\mapsto1$ &       7.0 & 0.375  $b_{\mathrm{inc}}^{2}$ \\
\hline
\rule[0pt]{0pt}{2.8ex}  $1\mapsto2$ &       13.5 & 0.75 $b_{\mathrm{inc}}^{2}$ \\
\hline
\rule[0pt]{0pt}{2.8ex}   $0\mapsto2$ &         21.0 & $b_{\mathrm{coh}}^{2} + 0.625 b_{\mathrm{inc}}^{2}$ \\
\hline
\rule[0pt]{0pt}{2.8ex}   $1\mapsto0$ &         7.0 & $0.75 b_{\mathrm{inc}}^{2}$ \\
\hline
\end{tabular}
\end{table}

In principle, $[b_{\rm{eff}}(Q)]^2$ is expressed by

\begin{equation}
\label{eq:beff}
\begin{aligned}
 \left[b_{\rm{eff}}(Q)\right]^{2}=& \,(1-c_p)\cdot\left[4\left(b_{\rm
coh}^2+\frac{5}{8}b_{\rm inc}^2\right)j_0^2(Qa_s/2)\right]\\
 & +(1-c_p)\cdot\left[\frac{9}{2}b_{\rm
inc}^2\left(j_1^2(Qa_s/2)\right)\right]\\
 & +c_p\cdot\left[3b_{\rm
inc}^2\left(2j_1^2(Qa_s/2)+3j_3^2(Qa_s/2)\right)\right]\\
& +c_p\cdot\left[(4b_{\rm coh}^2+b_{\rm
inc}^2)\left(j_0^2(Qa_s/2)+2j_2^2(Qa_s/2)\right)\right],\\
\end{aligned}
\end{equation}

where $j_i(Q\cdot a_s/2)$ are the spherical Bessel functions of
order i. The first term in Eq.\ \eqref{eq:beff} describes the
transition $J=0\mapsto 0$ (even-even), while the second term the
$J=0\mapsto 1$~(even-odd) transition. The third and fourth term
take the $J=1\mapsto 2$  (odd-even) and $J=1\mapsto 1$ (odd-odd)
transition into account. The parameter $a_s=0.74$~\AA~
 is the distance of the deuterons within the D$_2$ molecule
 \cite{Nielsen2}, while $c_p$ is the concentration of the
 molecules in the para state ($c_p+c_o=1$, $c_o$~is the concentration of molecules in the ortho state).
\\
\\
With the measured inelastic neutron cross section and Eq.\
\eqref{eq:csinc}, it is possible to determine the GDOS. The GDOS
for differently prepared sD$_2$ crystals with different ortho
concentrations are the main subject of this paper. Furthermore, a
methode of extracting the density of states of one-quasi-particle
excitations $G_1(E)$ from our data is presented. A detailed
analysis of the dynamical and static structure of solid deuterium
will be presented in a forthcoming publication.

\section{Experimental Details}
The experiments on inelastic neutron scattering were performed at
the Time-Of-Flight (TOF) spectrometer for thermal neutrons IN4
\cite{IN4} of the Institute Laue-Langevin (ILL), Grenoble, France
and at the cold TOF spectrometer TOFTOF \cite{TOFTOF} at
FRM II, Garching, Germany.\\
\\
The measurements at IN4 were carried out at two different
wavelengths of the incoming neutrons ($\lambda\simeq 2.2$~\AA~ and
$1.1$~\AA). The energy resolution for the two different
wavelengths had been determined to be $\Delta E_{2.2}=0.7$~meV and
$\Delta E_{1.1}=3.4$~meV, using the standard technique of
elastic neutron scattering on a vanadium sample.\\
\\
The IN4 spectrometer supplies a typical thermal neutron flux of
$5\cdot 10^{5}$ cm$^{-2}$ s$^{-1}$ on the sample. The beam size on
the sample is $\sim 2 \times$4~cm$^{2}$. The scattering angle
$2\theta$ varies between $13^{\circ}$ and $120^{\circ}$. The
beam divergence is $\Delta \theta=1^{\circ}$.\\
\\
The sample cell for the solid deuterium was designed and
constructed to fit into the standard "orange cryostat" of ILL
\cite{OK}. The sample cell is cooled down to the desired
temperature by using an exchange gas (helium) in the chamber which
surrounds the sample cell. The cooling inside the cryostat is done
by a liquid-nitrogen shield and a liquid-helium heat exchanger.
Neutron scattering on solid/liquid deuterium was performed at
temperatures between $T=3.5$~K and $T=23$~K. The temperature
stability was better than $0.1$~K.

The sD$_2$ is frozen inside the double-wall cylinder volume (see
Fig.\ \ref{fig:1}). The thickness of the sD$_2$ in the cell is
fixed to $\delta=2$~mm. The attenuation of the incoming thermal
neutrons ($E_0\sim 17$~meV) in the filled sample cell is $\sim
25\%$, and therefore multiple neutron scattering inside the sample
can not be neglected. The influence of multiple scattering will be
discussed in detail in the section "Results and discussion". The
sample cell is connected to the D$_2$
gas handling system via a $12$~mm$\times 1$~mm stainless steel pipe.\\

The gas handling system is especially designed to keep the
pressure of the D$_2$ gas flow into the sample cell
 quasi-independent of the temperature within the cell. This is done by
using needle valves for the gas flow. With this technique, it is
possible to follow a special path in the phase diagram  (see Fig.\
 \ref{fig:2}) of D$_2$. This can be either the path from the gas
via the liquid to the solid phase, or the path directly from the
gas to the solid phase (sublimation). The triple point of D$_2$ is
at [$T_{\mathrm{t}}=18.7$~K, $P_{\mathrm{t}}\sim$~180 mbar]. At
pressures below $P_t$, direct sublimation of D$_2$ is
 possible, while at pressures above $P_t$, the D$_2$ is liquefied before it becomes a solid. \\

The para-to-ortho converter \cite{PO3}, which accelerates the
conversion from the p-D$_2$ to o-D$_2$ state (Fig.\ \ref{fig:3}),
is attached to the gas handling system via a stainless-steel pipe.
This converter is an independent cryogenic system, cooled by a
cold finger. The main part of this system is a copper cup, filled
with a paramagnetic powder. This powder was extracted from an
OXISORB \cite{PO3} unit, which is normally a gas cleaning tool.
OXISORB contains paramagnetic chromium composition, which acts
inside the converter as a catalyst. The upper part of the copper
cup is closed by a sinter disc, which keeps the powder in the cup,
but is permeable to D$_2$ gas. This copper cup is mounted on the
finger of the cooling machine, and can be operated between
$T=10$~K and room temperature. During the filling mode, the cup is
fed with D$_2$ gas from the gas handling system. The D$_2$ is
 liquified inside the cup, and is kept there at the boiling point for
the necessary conversion time (hours). Keeping the liquid at the
boiling point shortens the conversion time considerably (factor 2) \cite{PO3}.\\
\\
The filling procedure of the sample cell was monitored with the
help of the IN4 spectrometer. Filling was stopped when the
scattered-neutron intensity reached its maximum and stayed
constant. This method was used for filling natural deuterium gas
from the dump
 and also for filling the converted deuterium gas from
the converter unit.\\
\\
The experiments at TOFTOF of FRM II were performed with an
 equipment similar to that used at IN4. These experiments
were performed before the IN4 experiment and served to optimize
the sample cell for the IN4 experiment. Sample cell and cooling
machine at the TOFTOF experiment were different from the setup at
IN4. The thickness of the sD$_2$ in the cell was bigger than that
($\delta=3$~mm) in the IN4 sample cell. The attenuation of the
incoming thermal neutrons ($E_0\sim 20$~meV, $\lambda\simeq
2.0$~\AA) in the filled sample cell is $\sim 30\%$, and therefore
multiple neutron scattering inside the sample can not be
neglected. The influence of multiple scattering will be discussed
in detail in the section "Results and discussion". Later, after
the IN4 experiment, a second experiment was done with sD$_2$ at
the TOFTOF, in order to improve the statistics of neutron data for
energy transfers larger than
$10$~meV. \\
The sample cell at TOFTOF is cooled by a closed
 cycle cold-head machine with He exchange gas. With this setup  temperatures down to
$T\sim$7~K were reached. The measurements at TOFTOF were  carried
out at three different wavelengths of the incoming neutrons
($\lambda\simeq 2.0$~\AA~, $\lambda\simeq 2.6$~\AA~  and
$\lambda\simeq 6.0$~\AA). The energy resolution for the different
wavelengths has been determined to $\Delta E_{2.0}=1.23$~meV,
$\Delta E_{2.6}=0.74$~meV and $\Delta E_{6.0}=0.07$~meV,
using the standard technique of elastic neutron scattering on a vanadium sample.\\
At TOFTOF the typical cold neutron flux  on the sample amounts to
$7\cdot 10^{4}$ cm$^{-2}$ s$^{-1}$. The beam size on the sample is
$\sim 3 \times 5$~cm$^{2}$.  The TOFTOF detector bank covers a
range of $2\theta =$ $7.5^{\circ}$ ... $140^{\circ}$. The
beam divergency  is $\Delta \theta\sim1^{\circ}$.\\

\section{Sample Preparation}
The structure and dynamics of sD$_2$ may be influenced by the way
of preparing the crystals \cite{Momose-1,Collins}. Therefore
samples of different character were prepared by different freezing
procedures. Our own experiments had shown differences in the
optical quality of D$_2$ for different freezing procedures (see
Fig. \ref{fig:4}).
\\
Freezing experiments with p-sH$_2$ and o-sD$_2$ films on MgF$_2$
coated sapphire plates \cite{Collins} showed very interesting
features concerning the microscopic and mesoscopic structure of
solid deuterium and hydrogen films. Below $T_d<6$~K (deposition
temperature) a mixture of hcp and fcc structures is possible in
the sD$_2$ film. The fcc component disappears if the temperature
is raised above $9.5$~K (annealing). The remaining hcp structure
is preserved, even if the temperature is lowered again below
 $9.5$~K. These observations were one motivation to look for
 a possible influence of the freezing procedure on the structure
 and dynamics of sD$_2$. The following freezing procedures were applied:\\
\\
\subparagraph{Liquid-Solid (\textbf{LS}):} D$_2$ is liquefied in
the sample cell at $T\sim$21~K. The filling is monitored by
neutron scattering measurements at IN4. After filling, the
temperature of the cooling controller is lowered to a temperature
slightly below $T=18.7$~K and kept constant, the sample starts to
freeze slowly. The sD$_2$ sample is then cooled down slowly to the
desired measurement temperature. Slow cooling of the sD$_2$ is
necessary to avoid sudden cracking of the solid deuterium.

\subparagraph{Liquid-Solid-Melt-Solid (\textbf{LSMS}):} An
alternative method of producing good samples of sD$_2$ is
remelting the solid at $T=18.7$~K. When the solid is melting, a
mixture of liquid and small pieces of solid deuterium crystals
will emerge. The character of this mixture can be monitored by
neutron-scattering measurements. The static structure factor
indicates the crossover from solid/liquid to complete liquid. The
basic idea of this method is to stop the melting shortly before
all solid pieces of sD$_2$ disappear. The remaining solid
fragments serve as crystallization seeds for a new solid-deuterium
sample.

\subparagraph{Gas-Solid (\textbf{GS}):} The re-sublimation of
deuterium from gas is done be keeping the sample-cell temperature
at $T\sim7-9$~K and maintaining a constant gas flow. With this
method it is possible to grow slowly a solid-deuterium sample in
the cell.
 How far the cell is filled with solid material is again monitored by
 neutron-scattering measurements.

\subparagraph{Turbo Solid (\textbf{TS}):} Fast freezing of sD$_2$
 from the liquid phase at a temperature below $T\sim10$~K is
leading to an opaque solid-deuterium sample (see Fig.
\ref{fig:4}c). The analysis of the static structure, extracted
from the neutron scattering data, shows an almost ideal powder
diffraction pattern for this "turbo solid". A detailed analysis of
the structure of sD$_2$ will be presented in a forthcoming paper.

\section{Results and Discussion}
The TOF data were corrected for the empty-cell measurements, and
normalized to the incoming flux and the vanadium standard. Based
on this data the GDOS were determined by the IN4 computer code
\textbf{gdos}. This code is implemented in the data analysis
package LAMP \cite{LAMP}, which is used for treating the data
obtained from neutron scattering experiments at ILL. The basic
method of this calculation is a sampling over a large Q range
(neutron energy loss side), using the incoherent approximation
\cite{Schober} and referring to Eq.\ \eqref{eq:csinc}.\\
The GDOS is calculated by:

\begin{equation}
\label{eq:GD}
\begin{aligned}
GDOS(\omega) = \omega \cdot\ \frac{S(Q(<2\theta>),\omega)} {Q^2
\cdot (n+1)}
\end{aligned}
\end{equation}
\\
The term $Q(<2\theta>)$ is a Q value for each energy transfer
$\hbar \omega$ considering an averaged scattering angle
$<2\theta>$. This simple approximation for the GDOS is commonly
not equal to the real density of states (DOS). In many cases the
agreement between the GDOS and DOS is remarkable. The GDOS
incorporates the one-phonon and multi-phonon excitations. In the
first order it is not so easy to separate the one-phonon part from
the multi-phonon part. Nevertheless the GDOS is a useful tool to
investigate the spectrum of excitations of the sample.

The experimental values of the GDOS for natural deuterium (o-D$_2$
 $66.7\%$) in the liquid and also in the solid phase for
different preparation methods of the crystals are shown in Fig.
\ref{fig:5}. The data for almost pure ($c_o=95\%$) o-D$_2$ are
shown in Fig. \ref{fig:6}.

The GDOS for differently prepared crystals at the same o-D$_2$
concentration does not show variations, beyond statistical
effects. It seems that the way of preparing the crystals does not
change the spectra of phonon excitations in sD$_2$. The main
difference between the GDOS for different o-D$_2$ concentrations
 is the strength of the rotational transition $J=0 \mapsto 1$.
Increasing the o-D$_2$ concentration enhances this transition.
 The GDOS shows for both o-D$_2$ concentrations a peak at
$E=$12..13 meV. Furthermore, at $c_o=66.7\%$ there is a hint for
the rotational transition $J=1 \mapsto 2$. This
excitation is not visible for $c_o=95\%$.\\

An example for the influence of the o-D$_2$ concentration $c_o$ on
the rotational transitions is shown in Fig.\ \ref{fig:7}. These
data are taken from the later TOFTOF measurements ($\lambda\simeq
2.0$~\AA). It is evident that increasing $c_o$ leads to a larger
cross section for the $J=0 \mapsto 1$ transition and also to a
decrease of the $J=1 \mapsto 0$ transition.

The main focus of the data analysis presented here is the
extraction of the density of states (DOS) $G_1(E)$ of phonons and
rotational transitions in solid D$_2$ ("quasi particle picture").
The basic idea is to compare the integral over the scattering
angle of the neutron scattering data $d\sigma/dE_f$ with a
calculated neutron cross section, using the incoherent
approximation developed by \textit{Turchin} \cite{Turchin} for a
cubic crystal with one atom in the primitive cell. With the aid of
this theory, it is possible to calculate the one-phonon and
multi-phonon contributions to the neutron cross section
$d\sigma/dE_f$, if $G_1(E)$ is known.\\
This approach is only valid, if the detected neutrons are
scattered neutrons  from all possible directions of the first
Brillouin zone in the crystal. This is normally the case, if one
has a powder sample of the crystals. The diffraction pattern of
fast frozen deuterium (TS, c$_p$=33.3\%, T=9.5K) shows an powder
like behavior, although some small texture effects are visible
(see Fig.\ \ref{fig:8}). A Rietveld fit on this diffraction
pattern gives the right values \cite{Mucker} for a and c of a hcp
solid deuterium crystal ($a=$3.596~+/-~0.005~\AA,
$c=$5.860~+/-~0.009~\AA). The positions of the different Bragg
peaks correspond to a hcp solid D$_2$ crystal.The fit-model
contains also the Debye-Waller factor, because at larger
scattering angles (higher Q-values) the scattering intensity is
significantly reduced. The fit result for the mean square
displacement $<u^2>=$0.208~+/-~0.011~$\AA^2$ is smaller compared
to value reported by \textit{Nielsen} \cite{Nielsen2}.
 These results induced the decision to use only data from TS
samples (second TOFTOF measurement) to determine the density of
states $G_1(E)$.

\begin{equation}
\label{eq:Sigma_s}
 \sigma_s(E_i\rightarrow
E_f)=\sum_{n=1}^\infty\sigma_n(E_i\rightarrow E_f),
\end{equation}

\begin{equation}
\label{eq:SigndE}
\frac{d\sigma_n}{dE_f}=\sigma_0(Q)\left(1+\frac{1}{\mu}\right)^2\sqrt{\frac{E_f}{E_i}}\frac{f_n(\epsilon)}{n!}F_n(E_i,E_f).
\end{equation}

\begin{equation}
\label{eq:Sig0}
\begin{aligned}
\sigma_0(Q) = 4\pi\cdot\left[b_{\rm{eff}}(Q)\right]^{2}
\end{aligned}
\end{equation}

The summation over the index \textsl{n} (number of quasi particles
 involved in the scattering) contains all higher-order multi-quasi-particle excitations like multiphonons. With ${E_i}$ and ${E_f}$
 are the energy of the incoming and outgoing neutron in the
scattering process, respectively, ${\mu}$ is the mass number of
the D$_2$ molecule (${\mu}=4$) and $\epsilon$ is the energy
transfer $(E=E_i-E_f)$ to the scattered neutron.

\begin{equation}
\label{eq:Ffn}
f_n(E)=\int_{-\infty}^{+\infty}f_{n-1}(E')f(E-E')dE',
\end{equation}

\begin{equation}
\label{eq:fE}
 f(E)=\frac{G_1(|E|)}{E\left(1-e^{-E/k_{\rm
B}T}\right)}\quad
\end{equation}

The function $f(E)$ (see (Eq.\ \eqref{eq:fE}) contains the density
of states $G_1(E)$, and it is possible to calculate the higher
order functions $f_n(E)$ with the help of Eq.\ \eqref{eq:Ffn}.

\begin{equation}
\label{eq:Fn}
F_n(E_i,E_f)=\frac{\mu}{4\sqrt{E_iE_f}}\int_{(\sqrt{E_i}-\sqrt{E_f})/\mu}^{(\sqrt{E_i}+\sqrt{E_f})/\mu}x^ne^{-x/\tau}dx
\end{equation}

\begin{equation}
\label{eq:tau}
\tau=\left(\int_0^\infty\frac{1}{E}\coth\left(\frac{E}{2k_{\rm
B}T}\right)G_1(E)dE\right)^{-1}
\end{equation}

In Eq.\ \eqref{eq:Fn} $F_n(E_i,E_f)$ is a kinematic factor,
resulting from the integration over all angles in the neutron
scattering process. The quantity $\tau$ is the average energy of
the excitations (mainly phonons) in sD$_2$.

The starting point of the determination of $G_1(E)$ is a
first-guess model for $G_1(E)$, which should contain the phonons
as well as the rotational transition $J \mapsto J'$. Following the
approach of \textit{Yu et al.} \cite{Yu} a model with seven
Gaussian functions is used to parameterize $G_1(E)$. The
contribution of the three-particle excitations ($d\sigma_3/dE_f$)
was approximated by an additional Gaussian function.

\begin{equation}
\label{eq:G1}
G_1(E)=\sum_{l=1}^7\alpha_l\frac{1}{\sqrt{2\pi}\sigma_l}e^{-\frac{1}{2}\frac{(E-\epsilon_l)^2}{(\sigma_l)^2}}
\end{equation}

Treating phonons and rotational transitions as independent
excitations may be not strictly correct in the calculation of the
multiphonon contribution, but an eventual correction would occur
at the far end of the energy spectrum above $E>14$~meV, which is
not crucial important here.

The neutron cross section $d\sigma /dE_f$ is calculated with Eqs.\
\eqref{eq:Sigma_s} - \eqref{eq:G1} and compared with the measured
$N_{0} \cdot d\sigma_{D} /dE_{f}$ ($N_{0}$ is a normalization
factor). The parameters $[\alpha_l,\sigma_l,\epsilon_l; l=1..7]$
are determined by minimizing the difference between measured and
calculated cross section.\\

Multiple scattering is in the case of the TOFTOF measurement not
negligible. The deuterium sample scatters approximately 25$\%$ of
the incoming beam. The scattering intensity is 50$\%$ elastic and
50$\%$ inelastic. The contribution of multiple scattering is a
flat falling background ($E>$3 meV) and peak-like at $E\sim$0 meV.
The multiple scattering was calculated (convolution of the
scattering model with itself) by using the experimental deduced
$G_1(E)$ and applying the \textit{Turchin} theory for one- and
two-phonon neutron cross section. The magnitude (~13$\%$) of the
multiple scattering was determined by using the work of
\textit{V.F. Sears} \cite{Sears} ("rule of thumb" \cite{Bacon}).
The flat falling background is smooth and has almost no texture.
This background was included in the fit model, and cross checked
afterwards with the convolution of $d\sigma /dE$ with itself. The
magnitude of multiple scattering, deduced from the fit must be in
the same order, as predicted by the "rule of thumb"-estimation.

The TOFTOF measurements (see Fig.\ \ref{fig:9})  clearly show the
onset of two-particle contributions above $E\sim5$~meV and also
one-particle excitations above $E=10$~meV. The TOFTOF data
indicates three-particle excitations above $E=12$~meV. The peak at
$E=5$~meV~ obviously originates from TA-phonons \cite{Nielsen1},
whereas the peak at $E\sim7.5$~meV is a signature of the
rotational transition $J=0 \mapsto 1$.

Figure \ref{fig:10} shows the result for the one-particle density
of states for solid D$_2$ at $c_o=66.7\%$ and $c_o=98\%$ (area
 normalized to 1). Part a) of Fig. \ref{fig:10} shows the
 convoluted $G_1(E)$, while in part b) $G_1(E)$ is de-convoluted
 with an Gaussian pseudo-resolution function with an
 $FWHM=$1.0~meV. This value was deduced from the $FWHM$ of the
 elastic peak of $d\sigma /dE$.

The influence of the ortho concentration $c_o$ on the density of
states $G_1(E)$ manifests itself (Fig.\ \ref{fig:10}) in the
expected enhancement of the peak at $E\sim7.5$~meV for larger
$c_o$. Increasing the number of ortho molecules in solid D$_2$
leads to a higher probability for $J=0 \mapsto 1$ transition. The
position of the $J=0 \mapsto 1$ transition depends slightly on the
concentration of para molecules $c_p$ in the sD$_2$. A larger
$c_p$ lowers the $J=0 \mapsto 1$  energy ($\delta E\sim0.3$~meV
for $c_p=33.3\%$). This energy shift can be explained by the
interaction between p-D$_2$ ($J=1$) and o-D$_2$ molecules ($J=0$)
\cite{Silv}.
\\
The DOS derived from our measurements differs remarkably from
earlier published results \cite{Yu}, which used the phonon
dispersion measurements of \textit{Nielsen} \cite{Nielsen2} to
calculate the DOS of solid D$_2$. The DOS published by \textit{Yu
et al.} \cite{Yu} contains pronounced peaks at $5$~meV and
$9$~meV, which were identified with the acoustic and optical
phonons in the hcp crystal. The $J=0 \mapsto 1$ transition was not
considered by \textit{Yu et al.} \cite{Yu}, and the DOS vanishes
above $10$~meV. The optical phonon contribution of the actually
measured DOS for $c_p$=33.3\% seems to be smeared-out in the
region of $8-12$~meV, and does not peak at $\sim9$~meV, as it was
published by \textit{Yu et al.}
 \cite{Yu}. This behavior was already observed in solid hydrogen by
 incoherent neutron scattering of \textit{Bickermann et al.} \cite{Bick} and \textit{Colognesi et al.} \cite{Colog}  .
\textit{Bickermann et al.}  explain the smearing-out of the
higher-energy optical phonon groups by the anharmonicity of the
quantum-crystal solid H$_2$. This anharmonicity leads to larger
phonon line widths at higher energies. The DOS of
\textit{Colognesi et al.} shows at $E\sim9-10$~meV a small bumb,
which could be identified as optic phonons in solid hydrogen. In
the case of almost pure o-D$_2$ ($c_p$=2\%) a peak at $E\sim9$~meV
in the DOS was found in the fit of our data. The amplitude of this
optical peak is smaller than as it was published by \textit{Yu et
al.}. The appearance of multiphonons above $5$~meV in our
measurements leads to
 considerable line widths of the phonon groups above this certain
energy. \textit{Nielsen's} DOS was determined by a pseudoharmonic
theory, which takes care of the renormalization of the phonon
energies, but does not consider the change of the phonon line
widths
\cite{Bick}.\\
A second new feature of the result for $G_1(E)$ presented here is
a clear peak-like structure at $E\sim12$~meV. The intensity of
this peak is more pronounced at higher $c_o$. This peak might be
caused by the excitation of a combination of the rotational
transition $J=0 \mapsto 1$ ($\sim 7.5$~meV) and phonons with
$\sim5$~meV energy. The DOS for phonons with $\sim5$~meV is very
large. A combined excitation of phonons and rotational transitions
of the D$_2$ molecules is certainly a result of the interaction of
the rotational transitions with the lattice of the solid D$_2$
crystal (phonon-rotation coupling) \cite{Kran}. A higher energy
phonon ($E>12$~meV) is able to excite a low energy phonon and also
the rotational transition $J=0 \mapsto 1$ through the
phonon-rotation coupling. This decay of higher-energy phonons was
maybe seen in our data.  This peak could be on the other hand
maybe an artefact, resulting from the incomplete separation of the
one-particle-excitations and two-particle-excitations in the DOS,
 and has to be further investigated.
\\
The DOS of solid D$_2$ with higher para concentration
($c_p=\sim33.3\%, c_o\sim66.7\%$) contains a peak structure (see
Fig.\ \ref{fig:10}) at $\sim14$~meV, which is originated by the
rotational transition $J=1 \mapsto 2$. This peak is not seen at
$c_p=2\%, c_o=98\%$. This effect is obvious, because the $J=1
\mapsto 2$ transition scales with the number of para molecules in
the solid D$_2$.\\
The result for $G_1(E)$ derived from our neutron scattering data
can be used to calculate the mean square displacement $<u^2>$.
This value ($<u^2>$=0.22~$\AA^2$ for $c_o\sim66.7\%$) is quite
close to the result from the Rietveld fit on the neutron
diffraction data of TS samples ($c_o\sim66.7\%$,
$<u^2>$=0.208~$\AA^2$), and smaller than the result reported by
\textit{Nielsen} \cite{Nielsen2}. In this context it is
interesting to mention the recent published work of \textit{Bafile
et al.} \cite{Bafile}, who reports on neutron diffraction
measurements of solid deuterium close to melting. The mean square
displacement increases to $<u^2>$=0.43~$\AA^2$ at $T\sim$18.7~K.
The conclusion of \textit{Bafile et al.} is the break down of the
pseudoharmonic approach for solid deuterium close to the melting
point.

\section{Conclusion}
In this manuscript, we reported investigations on the density of
states (DOS) of solid D$_2$ for different o-D$_2$ concentrations
with the powerful tool of neutron scattering. The experiments were
performed at time-of-flight instruments at ILL,
 Grenoble and at FRM II, Garching. The DOS was extracted from the
neutron data by applying the incoherent approximation. The results
for the GDOS (generalized density of states) and the DOS show
interesting details, which have not been published previously. One
feature is the appearance of the $J=0 \mapsto 1$ in the DOS. The
strength of this  excitation increases with the o-D$_2$
concentration $c_o$. The DOS of solid D$_2$ contains a strong
phonon signal at $5$~meV, but weak signals of optical phonons. Our
results for the DOS indicate a smearing-out of the higher energy
optical phonons for higher p-D$_2$ concentrations, which was also
reported for solid hydrogen with high o-H$_2$ concentrations. The
reason for this effect may be explained by the anharmonicity of
the solid D$_2$ crystal. Furthermore we see
 two-phonon processes at considerably low energies in our data
(onset at $5$~meV). Another new result is the appearance of a peak
at $\sim12$~meV in the DOS of solid D$_2$. This peak increases
with higher o-D$_2$ concentration, and could be explained by a
coupled excitation of a phonon and a rotational transition, or by
an possible incomplete separation of the one- and
two-particle-contribution. This effect has to be further
investigated. Our data of natural deuterium ($c_o\sim 66.7\%$) are
showing the rotational
transition $J=1 \mapsto 2$.\\
\\

\begin{acknowledgments}
This work was supported by the cluster of excellence "Origin and
Structure of the Universe" Exc 153 and by the
Maier-Leibnitz-Laboratorium (MLL) of Technische Universit\"at
M\"unchen (TUM) and Ludwig-Maximilians-Universit\"at (LMU). We
thank T. Deuschle, H. Ruhland and E. Karrer-M\"uller for their
help during the experiments and F.\ J.\ Hartmann for his helpful
comments and notes concerning our manuscript.
\end{acknowledgments}

\newpage 
\bibliography{INSP1D1}

\newpage

\begin{figure}[t]

\vspace{-2mm}

\begin{center}
\includegraphics[width=100mm]{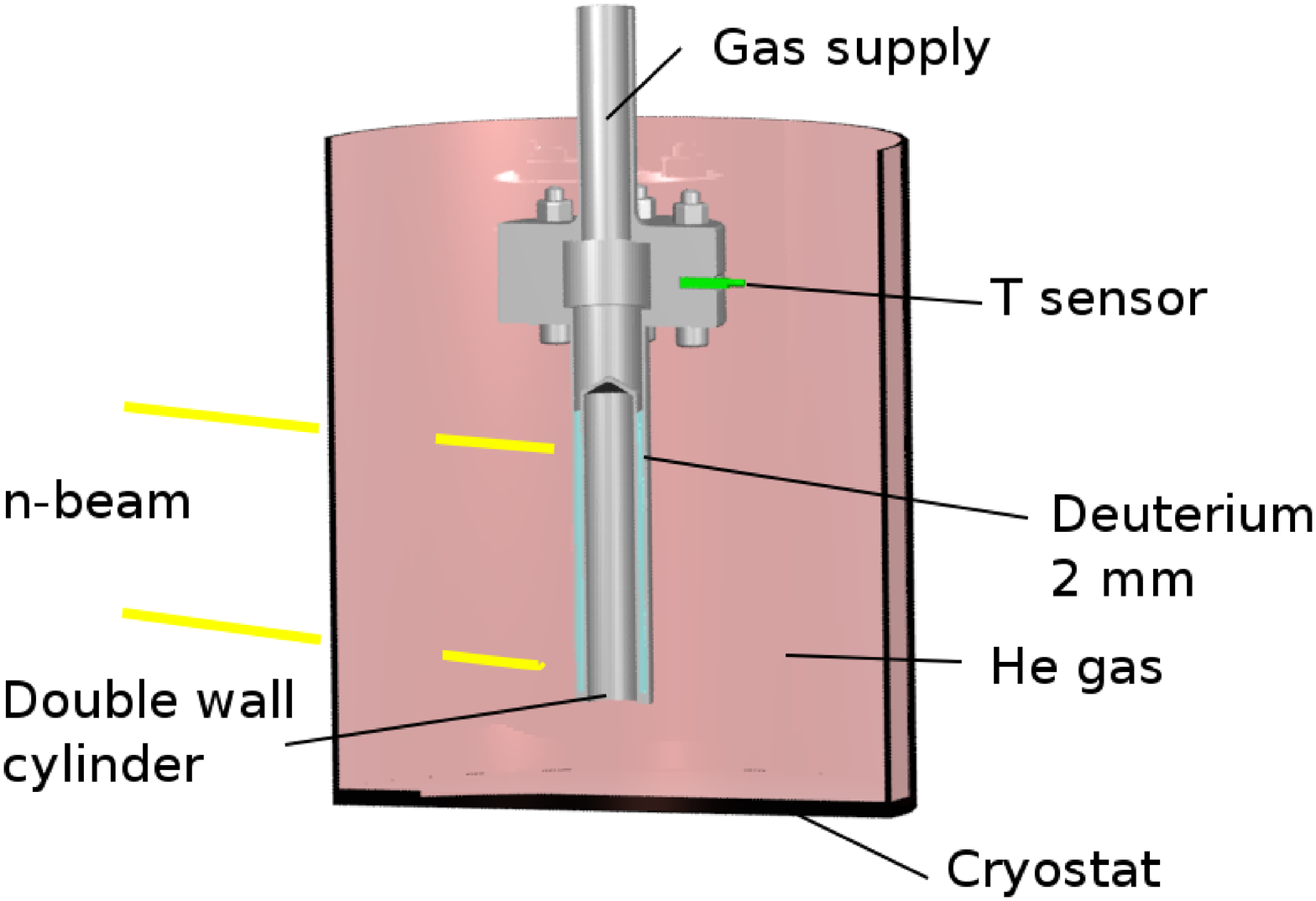}

\end{center}

\caption{~}{Sketch of the D$_2$ sample cell.}

\label{fig:1}

\vspace{5cm}

\end{figure}

\noindent
{\bf \large Figure captions:}\\
Fig.~\ref{fig:1}:\\
Sketch of the D$_2$ sample cell.  \\

\begin{figure}[t]

\vspace{-2mm}

\begin{center}
\includegraphics[width=100mm]{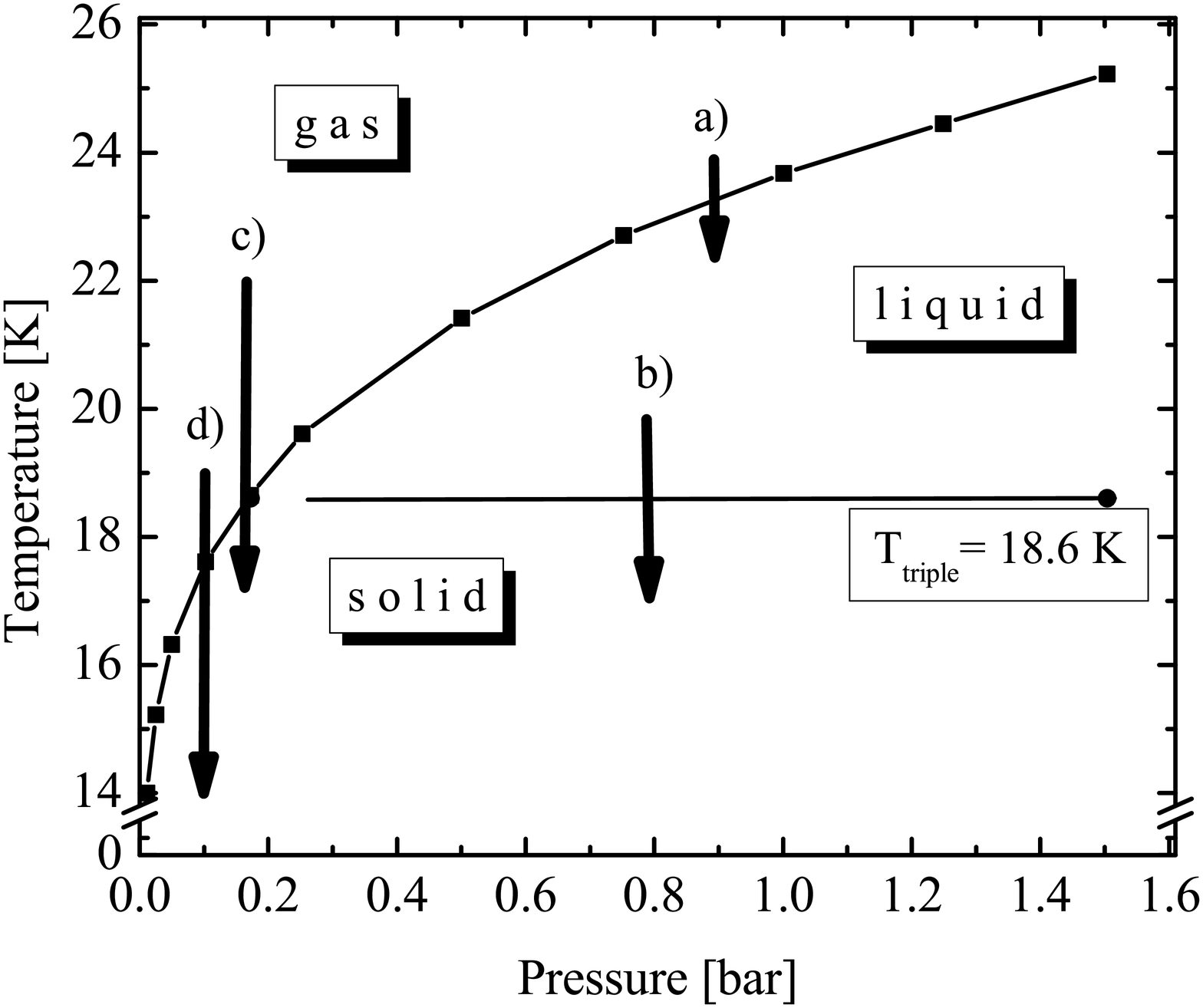}

\end{center}

\caption{~}{D$_2$ phase diagram with different process paths
studied:
 a) liquefaction from the gas, b) solidification from liquid, c)
sublimation at high temperature, d) direct condensation.}

\label{fig:2}

\vspace{5cm}

\end{figure}

\noindent
Fig.~\ref{fig:2}:\\
D$_2$ phase diagram with different process paths studied:
 a) liquefaction from the gas, b) solidification from liquid, c)
sublimation at high temperature, d) direct condensation.  \\

\begin{figure}[t]

\vspace{-2mm}

\begin{center}
\includegraphics[width=100mm]{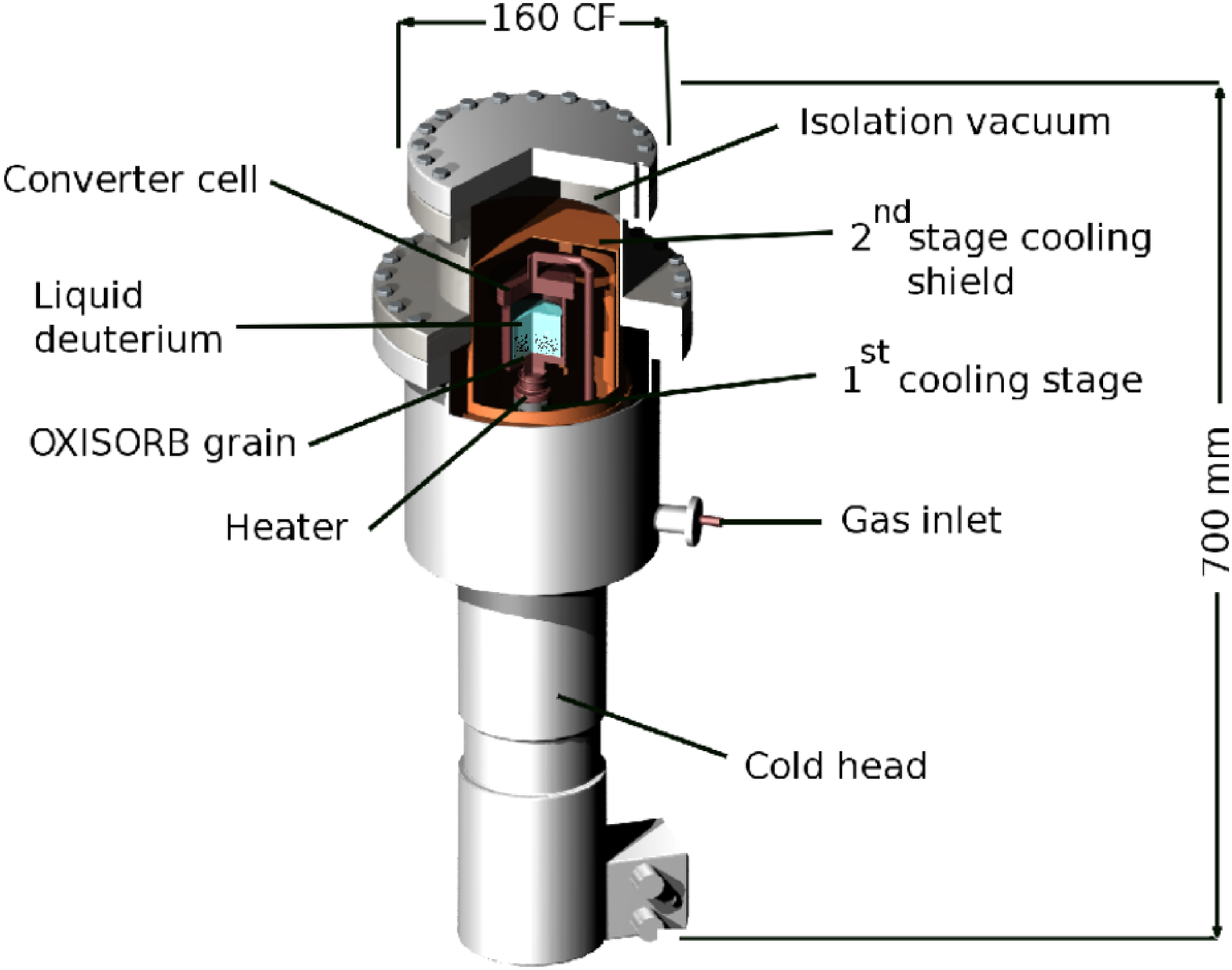}

\end{center}

\caption{~}{Sketch of the para-to-ortho converter \cite{PO3}. The
copper converter cup is filled with OXISORB grain, and mounted on
the cold-finger of the cooling machine.}

\label{fig:3}

\vspace{5cm}

\end{figure}

\noindent
Fig.~\ref{fig:3}:\\
Sketch of the para-to-ortho converter \cite{PO3}. The copper
converter cup is filled with OXISORB grain, and mounted on
the cold-finger of the cooling machine.  \\

\begin{figure}[t]

\vspace{-2mm}

\begin{center}
\includegraphics[width=100mm]{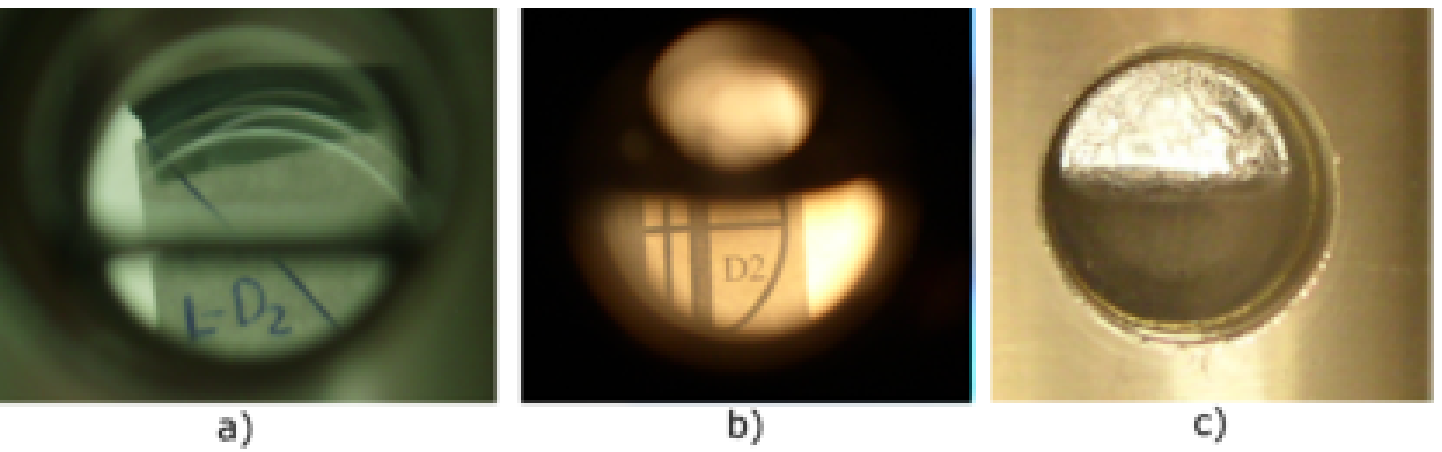}

\end{center}

\caption{~}{Pictures of solid deuterium samples for different
freezing procedures.\\ Crystals from the liquid phase are
transparent, if they are frozen slowly, while a fast freezing/cool
down (within minutes) leads to a non-transparent solid.
\\
a) solid deuterium slowly frozen from the liquid phase (LS),
\\
b) solid deuterium  slowly frozen from the gas phase (GS),
\\
c) solid deuterium fast freezing/cool down from the liquid phase
(TS).}

\label{fig:4}

\vspace{5cm}

\end{figure}

\noindent
Fig.~\ref{fig:4}:\\
Pictures of solid deuterium samples for different freezing
procedures.\\ Crystals from the liquid phase are transparent, if
they are frozen slowly, while a fast freezing/cool down (within
minutes) leads to a non-transparent solid.
\\
a) solid deuterium slowly frozen from the liquid phase (LS),
\\
b) solid deuterium  slowly frozen from the gas phase (GS),
\\
c) solid deuterium fast freezing/cool down from the liquid phase
(TS). \\

\begin{figure}[t]

\vspace{-2mm}

\begin{center}
\includegraphics[width=100mm]{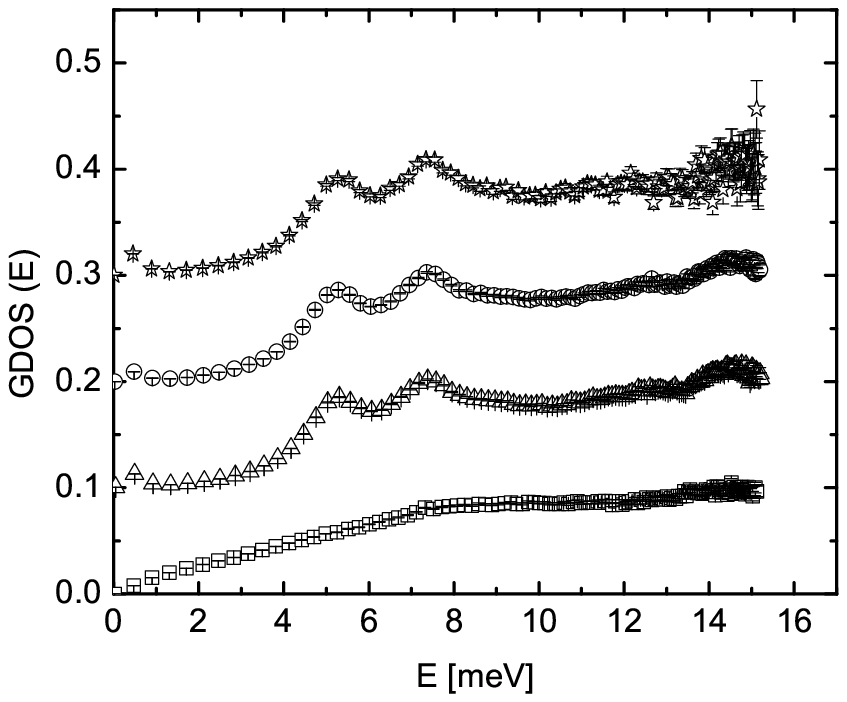}

\end{center}

\caption{~}{Generalized density of states (GDOS) of natural
deuterium ($c_o=66.7\%$):
\\ liquid D$_2$ ($\square$) at $T=$21~K, solid D$_2$
($\triangle$), rapidly frozen out from the liquid phase and fast
cooled down, solid D$_2$ ($\bigcirc$), slowly frozen out from the
liquid phase, solid D$_2$ ($\star$), slowly frozen out from the
gas phase. All solids have a temperature of $T=$4~K. Peaks are
convoluted with the IN4 energy resolution
curve: Normalized by $\int_0^\infty{GDOS(E)\cdot dE}=1$. \\
Each GDOS is separated by a shift of $0.1$.}

\label{fig:5}

\vspace{5cm}

\end{figure}

\noindent
Fig.~\ref{fig:5}:\\
Generalized density of states (GDOS) of natural deuterium
($c_o=66.7\%$):
\\ liquid D$_2$ ($\square$)~at $T=$21~K, solid D$_2$ ($\triangle$), rapidly frozen out from the liquid phase and
fast cooled down, solid D$_2$ ($\bigcirc$), slowly frozen out from
the liquid phase, solid D$_2$ ($\star$), slowly frozen out from
the gas phase. All solids have a temperature of $T=$4~K. Peaks are
convoluted with IN4 energy resolution curve:
Normalized by $\int_0^\infty{GDOS(E)\cdot dE}=1$. \\
Each GDOS is separated by a shift of $0.1$.

\begin{figure}[t]

\vspace{-2mm}

\begin{center}
\includegraphics[width=100mm]{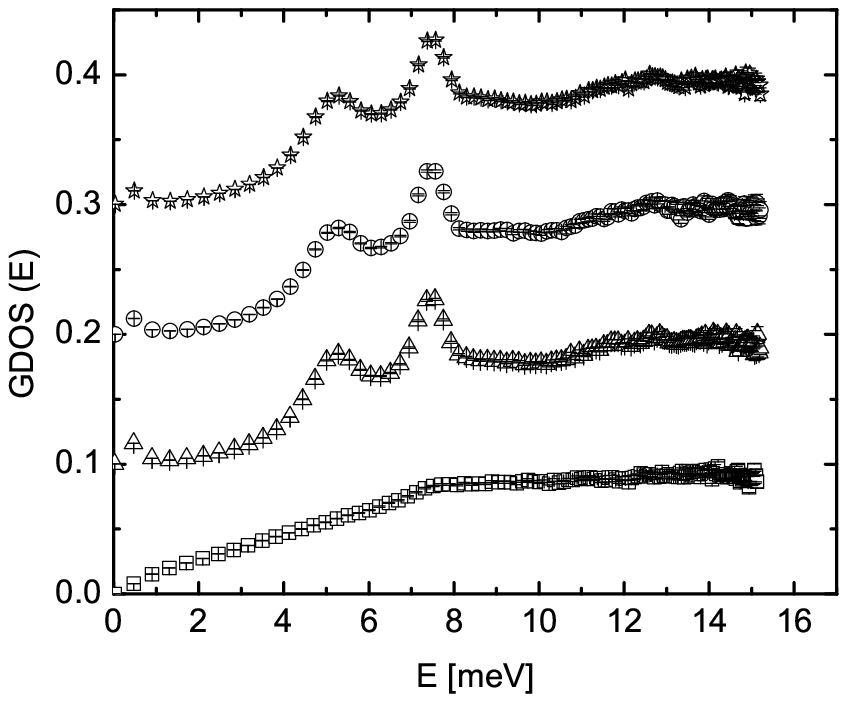}

\end{center}

\caption{~}{Generalized density of states (GDOS) of converted
deuterium ($c_o=95\%$):
\\ liquid D$_2$ ($\square$) at $T=$20~K, solid D$_2$ ($\triangle$), slowly frozen out from the liquid phase,
solid D$_2$ ($\bigcirc$), slowly frozen out from the gas phase,
solid D$_2$ ($\star$), frozen out at the melting point and slowly
cooled down. All solids have a temperature of $T=$4~K. Peaks are
convoluted with IN4 energy resolution
curve: Normalized by $\int_0^\infty{GDOS(E)\cdot dE}=1$.\\
Each GDOS is separated by a shift of $0.1$.}

\label{fig:6}

\vspace{5cm}

\end{figure}

\noindent
Fig.~\ref{fig:6}:\\
Generalized density of states (GDOS) of converted deuterium
($c_o=95\%$):
\\ liquid D$_2$ ($\square$)~at $T=$20~K, solid D$_2$ ($\triangle$), slowly frozen out from the liquid phase,
solid D$_2$ ($\bigcirc$), slowly frozen out from the gas phase,
solid D$_2$ ($\star$), frozen out at the melting point and slowly
cooled down. All solids have a temperature of $T=$4~K. Peaks are
convoluted with IN4 energy resolution
curve: Normalized by $\int_0^\infty{GDOS(E)\cdot dE}=1$. \\
Each GDOS is separated by a shift of $0.1$.

\begin{figure}[t]

\vspace{-2mm}

\begin{center}
\includegraphics[width=100mm]{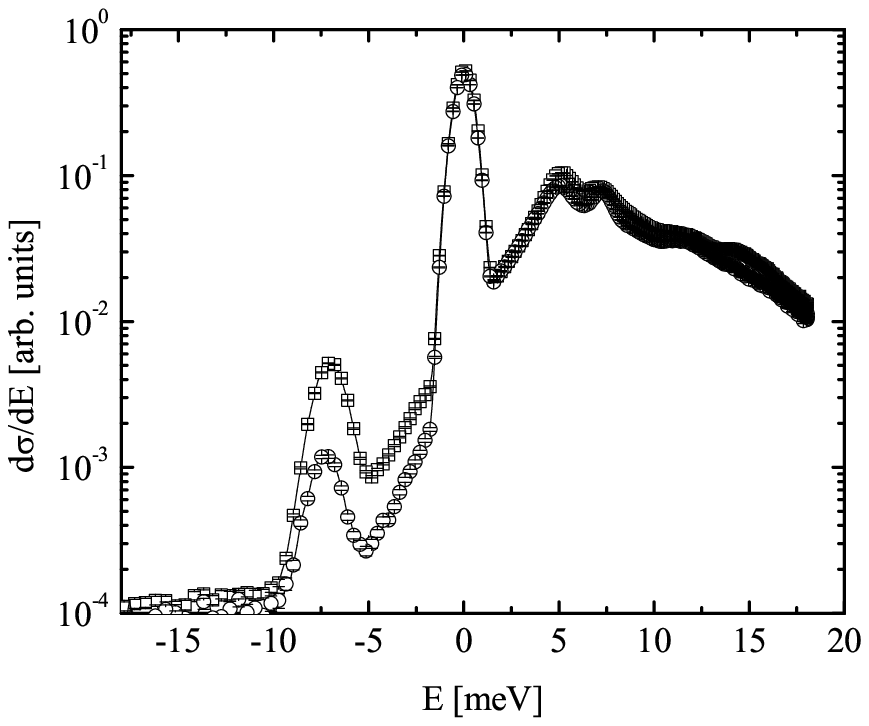}

\end{center}

\caption{~}{Dynamical neutron scattering cross section of solid
D$_2$ for $c_o=66.7\%$ ($\square$) and $c_o=98\%$ ($\bigcirc$) at
$T=7$~K. Data from the TOFTOF measurements.}

\label{fig:7}

\vspace{5cm}
\end{figure}

\noindent
Fig.~\ref{fig:7}:\\
Dynamical neutron scattering cross section of solid D$_2$ for
$c_o=66.7\%$ ($\square$) and $c_o=98\%$ ($\bigcirc$) at $T=7$~K.
Data from the TOFTOF measurements.\\

\begin{figure}[t]

\vspace{-2mm}

\begin{center}
\includegraphics[width=100mm]{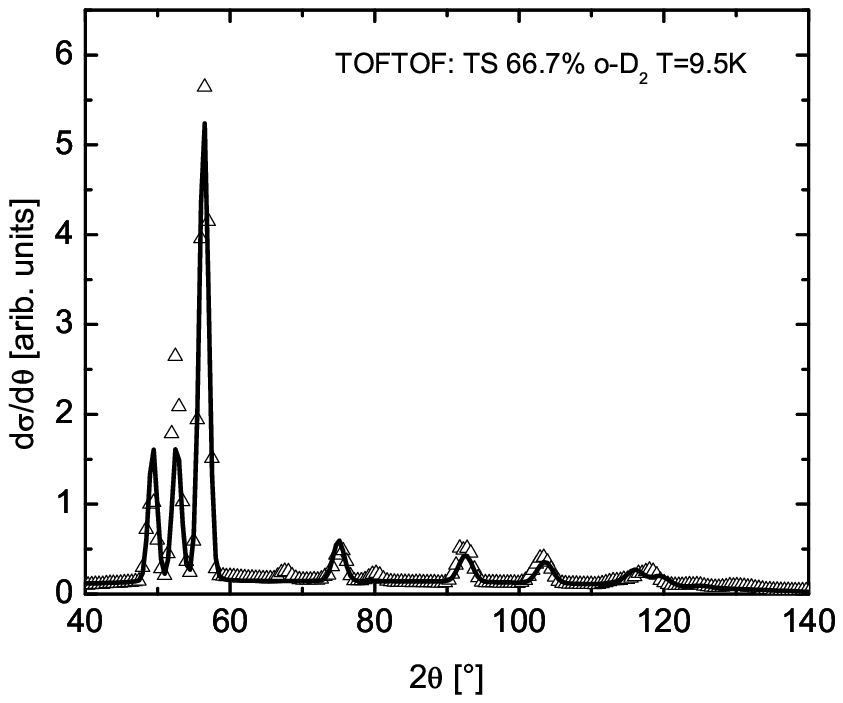}

\end{center}

\caption{~}{Neutron diffraction pattern data of fast frozen solid
D$_2$ (TS) for $c_o=66.7\%$ ($\triangle$) at $T=9.5$~K and
comparison with a Rietveld fit (solid line) for powder like hcp
solid D$_2$. Data from the TOFTOF measurements.}

\label{fig:8}

\vspace{5cm}
\end{figure}

\noindent
Fig.~\ref{fig:8}:\\
Neutron diffraction pattern data of fast frozen solid D$_2$ (TS)
for $c_o=66.7\%$ ($\triangle$) at $T=9.5$~K and comparison with a
Rietveld fit (solid line) for powder like hcp solid D$_2$. Data
from the
TOFTOF measurements.\\

\begin{figure}[t]

\vspace{-2mm}

\begin{center}
\begin{minipage}[t]{9 cm}
    \includegraphics[width=100mm]{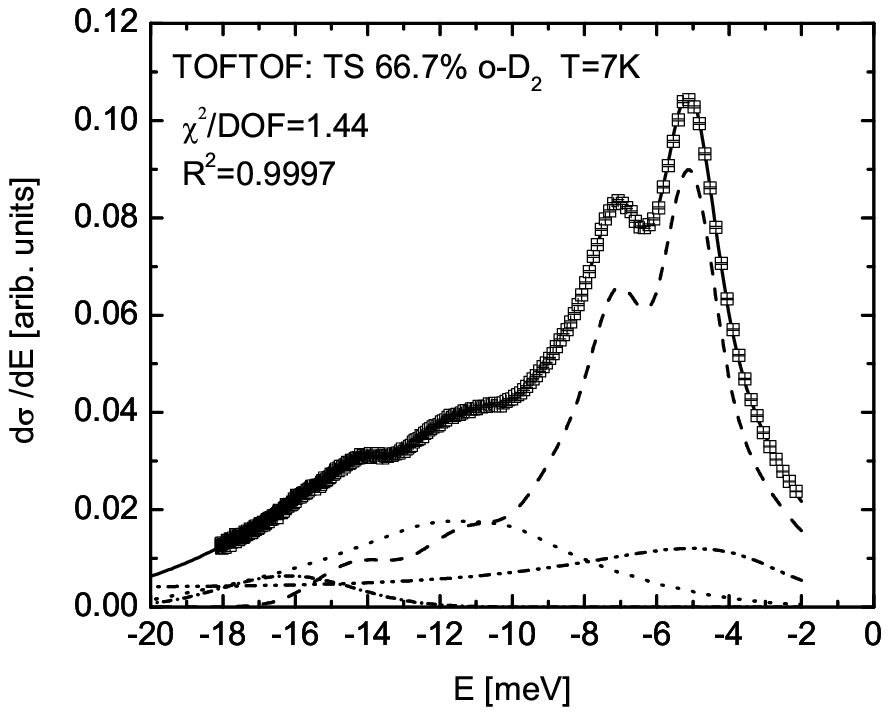}
  \end{minipage}
  \begin{minipage}[b]{9 cm}
    \includegraphics[width=100mm]{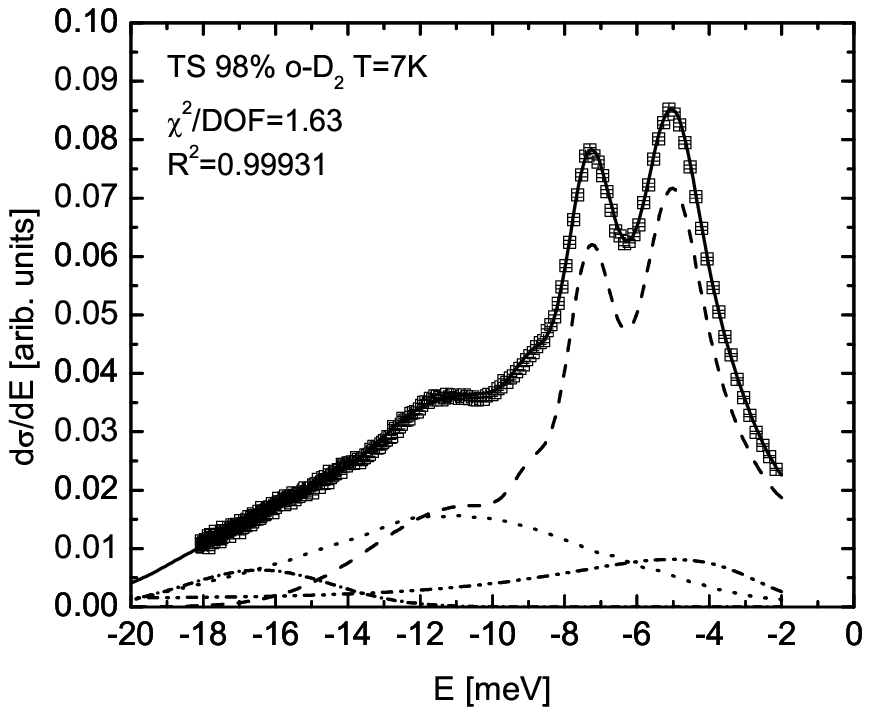}
  \end{minipage}

\end{center}

\caption{~}{Dynamic neutron scattering cross section of solid
D$_2$
for $c_o=66.7\%$ and $c_o=98\%$ at $T=7$~K. \\
Comparison of data with calculated neutron cross sections.\\ The
one-particle contribution is shown by the dashed line, the
two-particle contribution by the dotted line, and the
three-particle contribution by the dash-dotted line. Contribution
of multiple scattering is shown by the dot-dot-dash line. Data
from the TOFTOF measurements.}

\label{fig:9}

\vspace{5cm}

\end{figure}

\noindent
Fig.~\ref{fig:9}:\\
Dynamic neutron scattering cross section of solid D$_2$
for $c_o=66.7\%$ and $c_o=98\%$ at $T=7$~K. \\
Comparison of data with calculated neutron cross sections.\\ The
one-particle contribution is shown by the dashed line, the
two-particle contribution by the dotted line, and the
three-particle contribution by the dash-dotted line. Contribution
of multiple scattering is shown by the dot-dot-dash line. Data
from the
TOFTOF measurements.\\

\begin{figure}[t]

\vspace{-2mm}

\begin{center}

\begin{minipage}[t]{9 cm}
    \includegraphics[width=100mm]{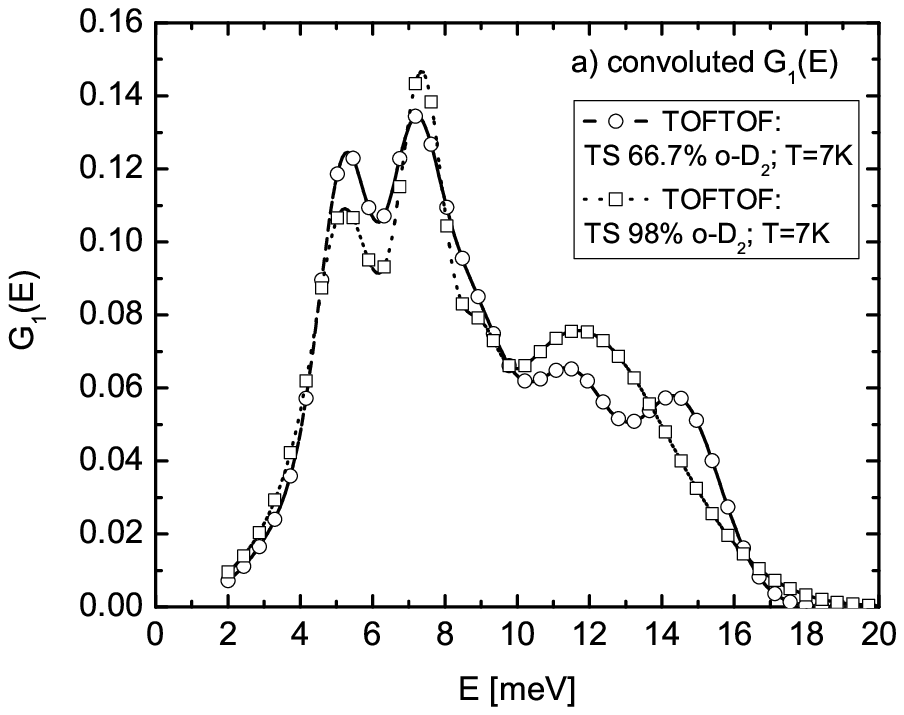}
  \end{minipage}
  \begin{minipage}[b]{9 cm}
    \includegraphics[width=100mm]{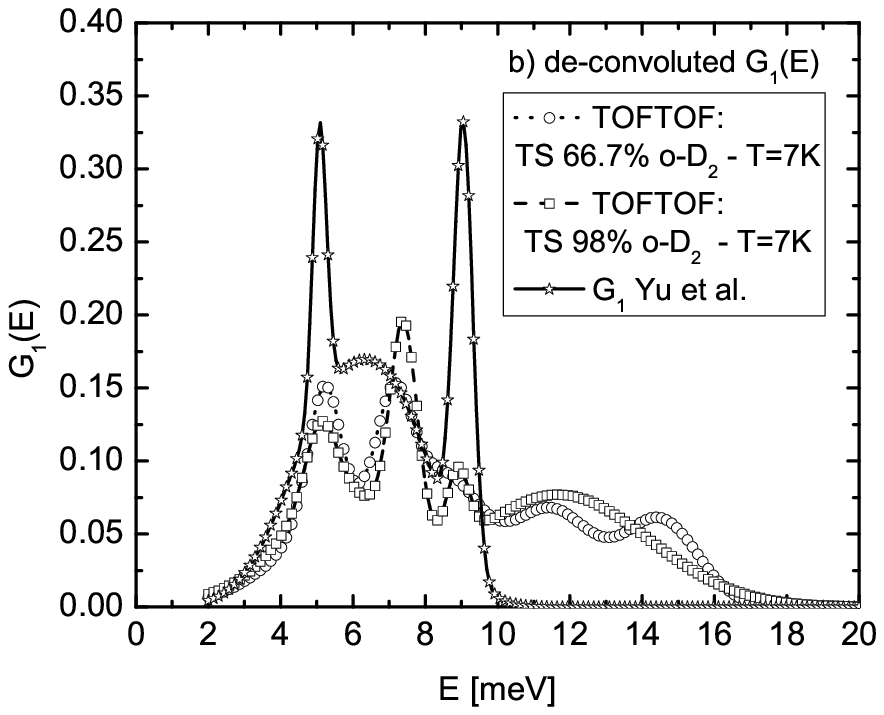}
  \end{minipage}

\end{center}

\caption{~}{a) One-particle density of states of solid D$_2$ for
$c_o=66.7\%$ ($\bigcirc$), $c_o=98\%$ ($\square$) at $T=7$~K.\\
b) Comparison of de-convoluted DOS with data ($\star$) from
\textit{Yu et al.}
\cite{Yu}.}\\
The DOS (part b)) are de-convoluted with the FWHM of the elastic
peak of $d\sigma /dE$.

\label{fig:10}

\vspace{5cm}

\end{figure}

\noindent
Fig.~\ref{fig:10}:\\
a) One-particle density of states of solid D$_2$ for
$c_o=66.7\%$ ($\bigcirc$), $c_o=98\%$ ($\square$) at $T=7$~K.\\
b) Comparison of de-convoluted DOS with data ($\star$) from
\textit{Yu et al.} \cite{Yu}.\\ The DOS (part b)) are
de-convoluted with the FWHM of the elastic peak of $d\sigma /dE$.
\\

\newpage

\newpage

\newpage

\newpage

\newpage

\newpage

\newpage

\newpage

\newpage

\newpage

\end{document}